\documentclass[aps,nofootinbib,showpacs]{revtex4-2}  
\usepackage{graphicx,subfigure}
\usepackage{float}
\usepackage{bm}
\usepackage{braket}
\usepackage{amsmath}
\usepackage{amssymb}
\usepackage{amscd}
\usepackage{latexsym}
\usepackage{slashed}
\usepackage{color}
\usepackage{graphicx}
\usepackage[normalem]{ulem}
\usepackage{color}
\definecolor{orange}{cmyk}{0,0.5,1,0}
\usepackage{verbatim}
\usepackage{rotating}
\usepackage{diagbox}
\usepackage{cancel}
\usepackage[utf8]{inputenc}
\usepackage{array}
\usepackage{hyperref}

\def\lsim{\raise0.3ex\hbox{$\;<$\kern-0.75em\raise-1.1ex\hbox{$\sim\;$}}}
\def\gsim{\raise0.3ex\hbox{$\;>$\kern-0.75em\raise-1.1ex\hbox{$\sim\;$}}}

\def\be{\begin{equation}}
\def\ee{\end{equation}}
\def\bea{\begin{eqnarray}}
\def\eea{\end{eqnarray}}
\def\nn{\nonumber}

\begin{document}
\title{The muon $g-2$ in an Aligned 2-Higgs Doublet Model with Right-Handed Neutrinos}

\author{Luigi Delle Rose}
\email[]{ldellerose@ifae.es}
\affiliation{\small Institut de Fisica d'Altes Energies (IFAE), The Barcelona Institute of Science and Technology, Campus UAB, 08193 Bellaterra (Barcelona), Spain}

\author{Shaaban Khalil}
\email[]{skhalil@zewailcity.edu.eg}
\affiliation{Center for Fundamental Physics, Zewail City of Science and Technology, 6 October City, Giza 12588, Egypt}

\author{Stefano Moretti}
\email[]{s.moretti@soton.ac.uk, stefano.moretti@physics.uu.se}
\affiliation{\small School of Physics and Astronomy, University of Southampton, Southampton, SO17 1BJ, United Kingdom}	
\affiliation{\small Department of Physics and Astronomy, Uppsala University, Box 516, SE-751 20 Uppsala, Sweden}	
	
\begin{abstract}
In this note, we show how one can attribute the anomaly currently present in the measurement of the anomalous magnetic moment of the muon ($a_\mu$) to the presence of  an underlying Aligned 2-Higgs Doublet Model (A2HDM) with Right-Handed (RH) neutrinos, $\nu_R$. The effects of the latter scenario are driven by one and two-loop topologies wherein a very light CP-odd neutral Higgs state ($A$) contributes significantly to $a_\mu$. Over the region of  parameter space of our new physics model which explains the aforementioned anomaly, also consistent with
the most recent measurements of the electron anomalous magnetic moment ($a_e$), wherein the charged Higgs state $H^\pm$ plays a
similar role,  we predict an almost background-free hallmark signature of it, due to $H^\pm A$ production followed by Higgs boson decays yielding multi-$\tau$ final states, which can already be searched for at the Large Hadron Collider (LHC). 
\end{abstract}
\maketitle

\section{Introduction}

It is a belief held by a large part of the particle physics community that the long-standing discrepancy between the Standard Model (SM) prediction for the muon anomalous magnetic moment and its experimental measurement is a clear hint of some New Physics (NP) Beyond the SM (BSM).
It all started with the 
 E821 experiment at Brookhaven National Laboratory (BNL) studying the precession of muons and antimuons in a constant external magnetic field as they circulated in a confining storage ring, which reported the following average value \cite{Muong-2:2006rrc}:
$a_\mu = 0.0011659209(6)$. New measurement techniques \cite{Semertzidis:1999kv,Farley:2003wt} enabled 
a more recent experiment at the Fermi National Laboratory (Fermilab), called `Muon $g-2$', using the E821 magnet, to already improve the accuracy of the $a_\mu$ value with the first batch of data obtained from March 2018 (the full run is expected to end in September 2022), as an {\sl ad interim} result released on 7 April  2021 yielded $a_\mu = 0.00116592040(54)$ \cite{Muong-2:2021ojo}, which, in combination with all other existing measurements, gives the most precise estimate to date as $a_\mu = 0.00116592061(41)$, exceeding the SM prediction by up to 4.2 standard deviations (depending on the assumptions made on the central value and error of SM prediction)\footnote{ The recent lattice determination of the hadronic vacuum polarisation contribution \cite{Borsanyi:2020mff} has questioned the significance of the anomaly. Nevertheless, its implications are still debated \cite{Crivellin:2020zul}.}. 
Finally, the  E34 experiment at the Japan Proton Accelerator Research Complex (J-PARC) plans to start its first run measuring $a_\mu$ in 2024 and is expected to improve accuracy further using new laser techniques \cite{Beer:2014ooa}. 

Such a sequence of measurements has generated growing interest in the particle physics community and several extensions of the SM have been proposed and analysed as possible origins of this result. Here, we study $a_\mu$ in a 2HDM with RH neutrinos and Aligned Yukawa couplings. In this class of models, one can account for the aforementioned deviations through two-loop effects generated by a light CP-odd neutral Higgs state ($A$) in combination with charged leptons. In fact, after improving the determination of the fine structure constant, it recently turned out that there is also a significant difference between the experimental result of the electron anomalous magnetic moment and the corresponding SM prediction.  In our BSM scenario, possible deviations in $a_e$ can be obtained through one-loop effects generated by the exchange of RH neutrinos and charged Higgs bosons and by exploiting the  lepton non-universality that naturally arises in RH neutrino models (we exploited RH neutrinos coupling predominantly to the first lepton family). 
Notably, the A2HDM with RH neutrinos can explain the two (potential) anomalies over the same region of parameter space. We refer to \cite{DelleRose:2020oaa} for more details.
Crucially, this phenomenology also requires the $H^\pm$ and $A$ states to be relatively light, so that their pair production processes have a sizeable cross section at the LHC, thereby enabling one to fingerprint this A2HDM with RH neutrinos in the years to come. 

According to the latest results, we have the following deviations in the anomalous magnetic moments of muon and electrons \cite{Davier:2017zfy,Keshavarzi:2020bfy,Parker:2018vye,Davier:2019can,Crivellin:2020zul,Colangelo:2020lcg,Davier:2010nc,Giudice:2012ms,Keshavarzi:2019abf,Aoyama:2020ynm}: 
\bea
\label{eq:g2mu}
\delta a_{\mu} &=& a_{\mu}^{\rm exp} - a_{\mu}^{\rm SM} = (251\pm 59) \times 10^{-11} \,, \nn \\
\delta a_{e} &=& a_{e}^{\rm exp} - a_{e}^{\rm SM} = (-87\pm 36) \times 10^{-14},
\eea
which indicate a $4.2 \sigma$ and $2.4\sigma$ discrepancy between theory and experiment, respectively \footnote{
The experimental value of $a_e$ is sensitive to the measurement of the fine-structure constant $\alpha$. The discrepancy quoted above is based on the measurement of caesium recoil by the Berkeley experiment \cite{Parker:2018vye}
Recently, a different experiment \cite{Morel:2020dww} reported a result for $\alpha$ that implies a $+1.6 \sigma$ deviation from the SM. The two experiments appear to be inconsistent with each other.
}. 
Here, it is crucial to notice that the sign of the two anomalies is different, prompting the realisation that they cannot be resolved simultaneously with the
same NP contribution, unless it violates Lepton Flavour Universality (LFU) in a very peculiar way, see, e.g., Refs. [13-37] in \cite{DelleRose:2020oaa}.

\section{The A2HDM with RH neutrinos}

We refer to \cite{DelleRose:2019ukt,DelleRose:2020oaa} for the details on the model while here we only discuss the most relevant features.
As for the neutrino sector, we simply note that  the heavy $\nu_R$ states generate the light neutrino masses through a seesaw mechanism. 
Concerning the 2HDM Higgs sector, we note that, other than requiring the usual $Z_2$ symmetry, the potentially dangerous tree-level Flavour Changing Neutral Currents (FCNCs) can also be tamed by implementing an alignment in flavour space. This implies that  
the two Yukawa matrices that couple to the same RH quark or lepton are proportional to each others with a coefficient $\zeta_f$. We assume a real $\zeta_f$ to avoid extra sources of CP-violation and  the alignment also in the neutrino sector (even though it is not strictly required by the absence of FCNCs). Because of such an alignment, all the couplings of the scalar fields to fermions are proportional to the corresponding mass matrices. As such, this 2HDM realisation is different from the standard four Types \cite{Gunion:1989we,Gunion:1992hs,Branco:2011iw}, wherein the interactions are fixed by $\tan\beta$, the ratio of the Vacuum Expectation Values (VEVs) of the two Higgs doublets.

The Yukawa interactions of the (pseudo)scalar states are given by
\bea
- \mathcal L_Y &=&  \frac{\sqrt{2}}{v} \bigg[   \bar u  ( - \zeta_u \, m_u \, V_{ud} \, P_L + \zeta_d \, V_{ud} \, m_d \, P_R  ) d   
+ \bar \nu_l  ( - \zeta_\nu \, m_{\nu_l} \, U^\dag_{L l}  \,  P_L   + \zeta_\ell  \,  U^\dag_{L l}  \, m_\ell \, P_R )  \ell       \nn \\
&+&  \bar \nu_h  ( - \zeta_\nu \, m_{\nu_h} \, U^\dag_{L h}  \,  P_L   + \zeta_\ell  \,  U^\dag_{L h}  \, m_\ell \, P_R )  \ell   \bigg]  H^+   \nn \\
&+& \frac{1}{v} \sum_{\phi=h,H,A} \sum_{f=u,d,\ell} \xi_f^\phi \, \phi \, \bar f  \, m_f \, P_R \,  f      \nn \\
&+& \frac{1}{v} \sum_{\phi=h,H,A} \xi_\nu^\phi \, \phi (\bar \nu_l \, U_{Ll}^\dag  + \bar \nu_h \, U_{Lh}^\dag) P_R (U_{Ll} \, m_{\nu_l} \, \nu_l^c + U_{Lh} \, m_{\nu_h} \, \nu_h^c) + \textrm{h.c.}, 
\eea
where $\zeta_f$ are the proportionality coefficients introduced above, $V_{ud}$ is the usual Cabibbo-Kobayashi-Maskawa (CKM) matrix and $P_{L,R}$ are the chiral projectors. The couplings of the neutral Higgs states to the fermions are given by 
\bea
\xi_{u, \nu}^\phi = \mathcal R_{i1} + ( \mathcal R_{i2} - i  \mathcal R_{i3} ) \zeta_u^*   \,, \qquad
\xi_{d,\ell}^\phi = \mathcal R_{i1} + ( \mathcal R_{i2} + i  \mathcal R_{i3} ) \zeta_{d,\ell}, 
\eea
with the matrix $\mathcal R$ diagonalising the scalar mass matrix. The $U_{L l}$ and $U_{L h}$ are the components of the mixing matrix in the neutrino sector,
\bea
\left( \begin{array}{c} \nu_L \\ \nu_R^c \end{array} \right) = U \left( \begin{array}{c} \nu_l \\ \nu_h \end{array} \right) \equiv \left( \begin{array}{cc} U_{Ll} & U_{Lh} \\ U_{R^c l} & U_{R^c h} \end{array} \right) \left( \begin{array}{c} \nu_l \\ \nu_h \end{array} \right) \,.
\eea
Finally, the  neutral and charged gauge boson interactions of the neutrinos, respectively, are:
\bea
 \mathcal L_Z &=& \frac{g}{2 \cos \theta_W}  (\bar \nu_l \, U_{Ll}^\dag + \bar \nu_h \, U_{Lh}^\dag) \gamma^\mu  (U_{Ll} \, \nu_l  +  U_{Lh} \, \nu_h )  Z_\mu, \nn \\
 \mathcal L_W &=& - \frac{g}{\sqrt{2}} \left[  (\bar \nu_l \, U^\dag_{L l} + \bar \nu_h \, U^\dag_{L h}) \gamma^\mu P_L \, \ell \right] W^{+}_\mu  + \textrm{h.c.}
\eea

\section{The $\mu$ and $e$ anomalous magnetic moments}

\begin{figure}
\includegraphics[width=5 cm]{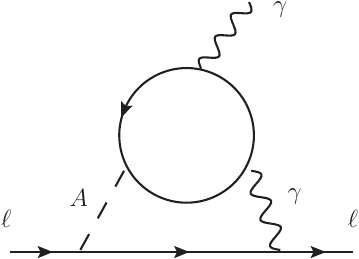}\qquad \qquad
\includegraphics[width=5 cm]{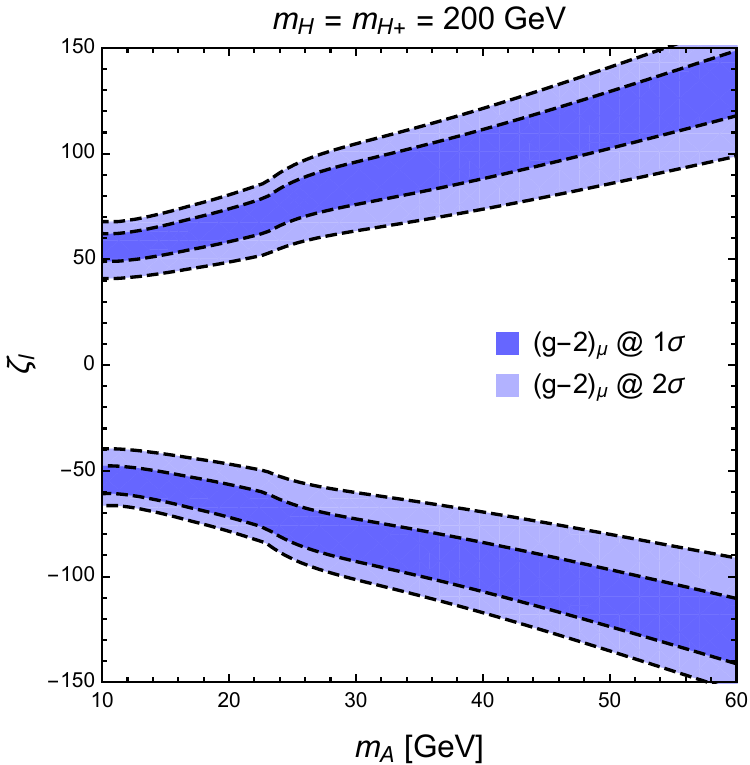} 
\caption{(Left) The contribution of the pseudoscalar $A$ through the two-loop Barr-Zee diagram. (Right) The 1 and $2\sigma$ regions of the anomalous magnetic moment of the muon in the parameter space of $m_A$ and $\zeta_\ell$. For the sake of definiteness, the mass of the charged Higgs boson has been chosen as $m_{H^\pm} = 200$ GeV.}
\label{AMM-muon}
\end{figure}

As customary in many 2HDM realisations, the solution of the $a_\mu$ anomaly relies upon a light pseudoscalar state $A$ contributing through the two-loop Barr-Zee diagrams, see left panel in Figure~\ref{AMM-muon}.
The explanation  is particularly simple in the `lepton-specific' 2HDM scenario in which the couplings of the $A$ and  $H^\pm$ scalars to the leptons can be enhanced. At the same time, those to the quarks are suppressed thus avoiding the strong constraint from the perturbativity of the top-quark Yukawa coupling (typical of Type-I and -III) or the severe bounds imposed  by flavour physics (such as in Type-II). All these issues can be naturally addressed in the A2HDM since the couplings to leptons and quarks are disentangled, namely $\zeta_\ell$ can be enhanched independently of $\zeta_u$ and $\zeta_d$.
Moreover, it is worth emphasising that a simultaneous explanation of the $a_\mu$ anomaly and a possible deviation in the $a_e$ cannot be neither achieved  in scenarios with a discrete $Z_2$ symmetry nor in the pure aligned 2HDM, since the contributions to the magnetic moments have a fixed sign originating from the same $\zeta_\ell$. Here, instead, the degeneracy will be broken by exploiting the LFU breaking that naturally arises in the RH neutrino sector\footnote{See \cite{Botella:2020xzf,Penuelas:2017ikk,Botella:2018gzy} for other solutions.}.

The results of our analysis are depicted in the right plot in Figure~\ref{AMM-muon},  which shows the regions in which the predicted $a_\mu$ is within 1 and $2\sigma$ around the measured central value
\bea
\delta a_{\mu} &=& a_{\mu}^{\rm exp} - a_{\mu}^{\rm SM} = (251\pm 59) \times 10^{-11} \,. 
\eea
These are projected onto the most relevant parameter space given by $m_A$ and $\zeta_\ell$ while the charged Higgs boson mass is fixed to a reference value of $m_{H^\pm} = 200$ GeV. Different choices of $m_{H^\pm}$ would only slightly modify the contours shown in the plot. Details of the scan of the parameter space can be found in \cite{DelleRose:2020oaa}, here we just mention the results complying with all experimental and theoretical bounds: namely, direct LEP searches, EW precision observables, flavour physics, Higgs direct and indirect searches at the LHC, LFU violation as well as vacuum stability, perturbativity and unitarity, respectively.

\begin{figure}
\includegraphics[width=5 cm]{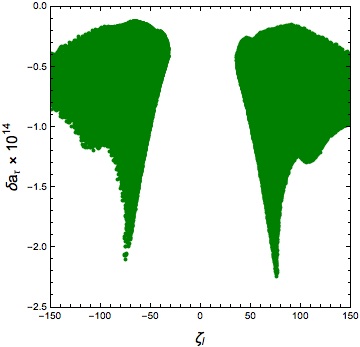}
\caption{The anomalous magnetic moment of the $\tau$ lepton as function of $\zeta_\ell$.}
\label{AMM-tau}
\end{figure}

As an additional prediction of the model, we show in Figure~\ref{AMM-tau} the maximum allowed BSM correction to the AMM of the $\tau$ lepton as a function of $\zeta_\ell$. The region of the parameter space depicted in the plot also reproduces the AMMs of the muon and the electron as in Eq.~\ref{eq:g2mu}. Among all the constraints, the most relevant one is the LFV process $\tau \to e \gamma$ which sets a strong bound on the product of mixings $(U_{Lh})_{\tau \nu_h} (U_{Lh})_{e \nu_h}$. In Figure~\ref{AMM-tau} we show the $\delta a_\tau$ corresponding to the maximum value of the mixing matrix element $(U_{Lh})_{\tau \nu_h}$ compatible with the aforementioned constraint. It is worth mentioning that if one does not insist in reproducing $\delta a_e$ in Eq.~\ref{eq:g2mu}, then the predicted value of the AMM of the $\tau$ could be further increased.
Although the current precision is far from probing these corrections, the best measurement at 95\% CL is $- 0.052 < a_\tau < 0.013$, the possibility to improve this determination in the future would be very interesting for constraining BSM scenarios and could be directly correlated to the collider predictions discussed in the next section.

\section{Collider signals}

Within the parameter space defined above and characterised by a light $A$ and $H^\pm$ with leptophilic interactions, the main decay modes for the BSM (pseudo)scalars are summarised by the following list:
\begin{itemize}
\item $A \to \tau \tau$
\item $H^\pm \to \tau^\pm \nu, H^\pm \to A W^\pm$
\item $H \to \tau \tau, H \to A Z$
\end{itemize}
The $A$ state decays at tree-level to $\tau$ pairs with Branching Ratio (BR) close to $100\%$. Furthermore,  for the charged Higgs boson, a leptonic decay mode is determined by the coupling $g_\ell = \zeta_\ell \, m_\tau / m_{H^\pm}$ while the strength of the $H^\pm \to A W^\pm$ channel is completely fixed by the $SU(2)_L$  coupling constant. In fact, the decay modes of the heavy neutral scalar state, $H$, share the same properties of those of the charged Higgs one. For large $m_{H^\pm}, m_H$ and neglecting small deviations from $\sin(\beta-\alpha) = 1$, the BRs of the $H^\pm$ and $H$ can be approximated by
\bea
&& \textrm{BR}(H^\pm \to A W^\pm) = \textrm{BR}(H \to A Z) = \frac{1}{1 + 2 g_\ell^2} \,, \nn \\  && \textrm{BR}(H^\pm \to \tau^\pm \nu) = \textrm{BR}(H \to \tau \tau) = \frac{2 g_\ell^2}{1 + 2 g_\ell^2}   \,.
\eea

\begin{figure}[H]
\centering
\includegraphics[width=4 cm]{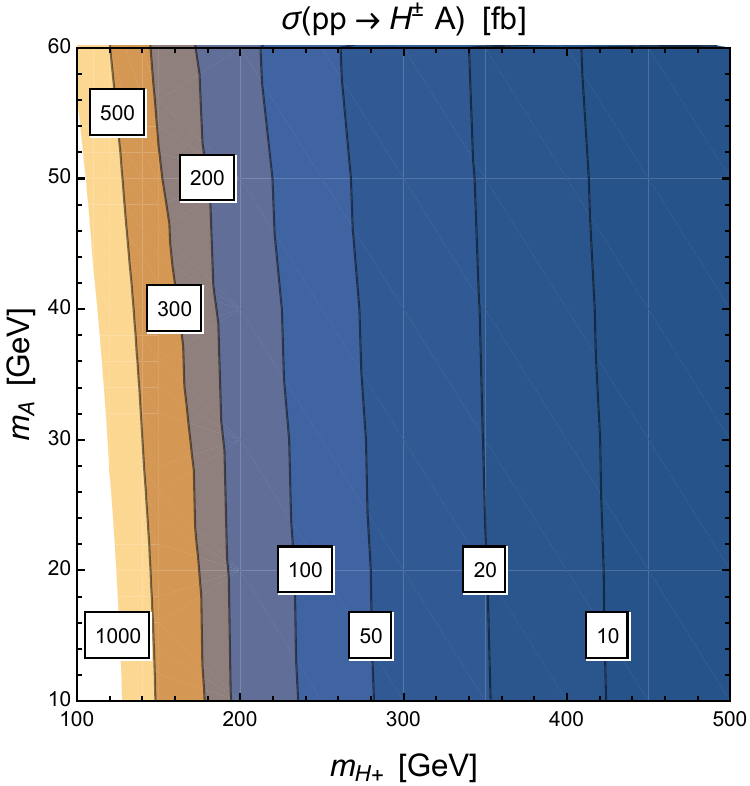}
\includegraphics[width=4 cm]{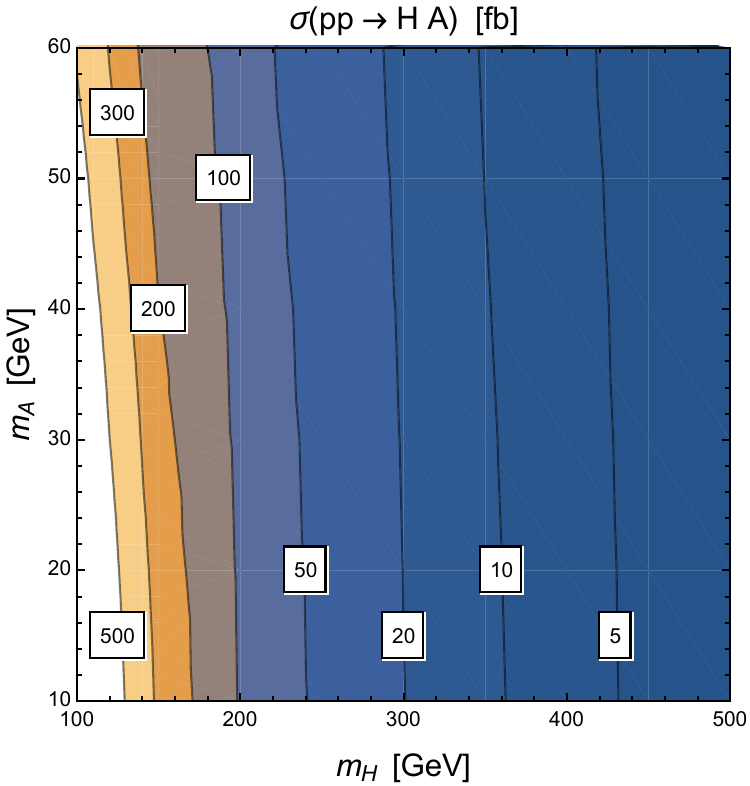}
\includegraphics[width=4 cm]{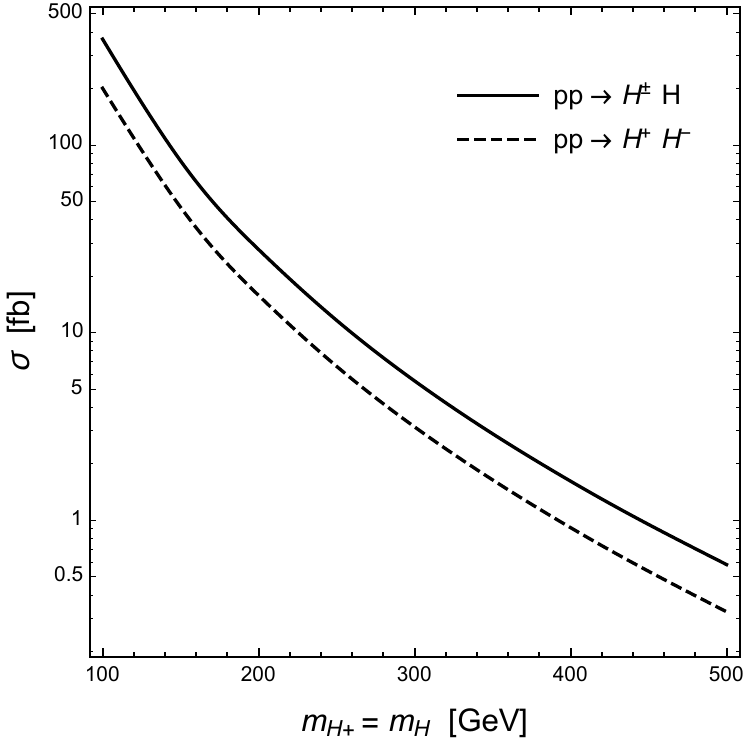}
\caption{The production cross sections of the extra Higgs boson pairs at the LHC with $\sqrt s =13$ TeV as functions of $m_A$ and $m_{H^\pm} = m_H$. \label{fig:xs}}
\end{figure}

The main production modes of these BSM (pseudo)scalar states proceed through Electro-Weak (EW) interactions since the couplings to the quarks are suppressed. Therefore, the relevant channels are
\bea
pp \to H^\pm A \,, \qquad pp \to H A \,, \qquad  pp \to H^\pm H \,, \qquad  pp \to H^+ H^-,
\eea
and the corresponding cross sections, shown in Figure~\ref{fig:xs}, depend only on the masses of the produced Higgs bosons. 

The main signatures resulting from the  production  and decay channels discussed above are characterised by final states with several $\tau$ leptons:
\bea
3 \tau + \slashed{E}_T, \qquad 4 \tau + W^\pm, \qquad 4 \tau  , \qquad 4 \tau + Z,
\eea
where the first two mainly arise from $H^\pm A$ production (with a subleading contamination from $H^\pm H$) while the other two stem from $HA$ production. 

In order to get a feel of the discovery potential of these channels at the LHC, with cross sections that can easily reach $10 - 10^2$ fb, here we also provide an estimate of the inclusive cross sections of the relevant SM backgrounds:
\bea
& \sigma_\textrm{SM}(Z W^\pm \to 3 \tau + \slashed{E}_T) \simeq 94 \, \textrm{fb}, \qquad 
& \sigma_\textrm{SM}(Z Z W^\pm \to 4 \tau + W^\pm) \simeq 3.2 \times 10^{-2} \, \textrm{fb}, \nn \\ 
& \sigma_\textrm{SM}(Z Z \to 4 \tau) \simeq 11 \, \textrm{fb}, \qquad 
& \sigma_\textrm{SM}(Z Z Z \to 4 \tau + Z ) \simeq 1.1 \times 10^{-2} \, \textrm{fb} \,.
\eea
The signal-to-background ratios in the above channels can therefore be of $O(1)$ or higher, thereby motivating future analyses by ATLAS and CMS already at Run 2. 

\section{Conclusions}

The current $a_\mu$ measurement is amongst the most precise ones experimentally, and so is also that of $a_e$, thereby probing not only the structure of the SM but also the possibility of BSM extensions.
The anomaly appearing in the $g-2$ mesurement of the muon and/or electron finds a natural explanation in the A2HDM supplemented by RH  neutrinos. In such a scenario, the contribution of a very light CP-odd neutral Higgs state interacting with leptons and the interplay of a charged Higgs boson with heavy neutrinos at the EW scale  can easily explain the deviations from the SM predictions with one and two-loop corrections in either or both observables. Rather crucially,  such a dynamics also predicts new signals at the LHC from  $H^\pm A$ production yielding multi-$\tau$ final states, which can even be background free (in some BSM region of parameter space) and thus accessible already with current LHC data. 

\bibliographystyle{apsrev4-1}
\bibliography{draftbib}

\end{document}